\documentclass[11pt]{article}

\usepackage{acl}
\usepackage{graphicx}
\usepackage{tabularx}
\usepackage{caption}
\usepackage{subcaption}
\usepackage{xurl}

\usepackage[utf8]{inputenc} % allow utf-8 input
\usepackage[T1]{fontenc}    % use 8-bit T1 fonts
\usepackage{hyperref}       % hyperlinks
\usepackage{url}            % simple URL typesetting
\usepackage{booktabs}       % professional-quality tables
\usepackage{amsfonts}       % blackboard math symbols
\usepackage{nicefrac}       % compact symbols for 1/2, etc.
\usepackage{microtype}      % microtypography
\usepackage{xcolor}         % colors

\usepackage{enumitem}
\usepackage{algorithm}
\usepackage[noend]{algpseudocode}
\usepackage{amsmath}
\usepackage{amssymb}
\usepackage{caption}
\usepackage{makecell}
\usepackage{multirow}
\usepackage{wrapfig}
\usepackage{graphicx}
\usepackage{subcaption}
\usepackage{pifont}

\usepackage{algorithm}
\usepackage[noend]{algpseudocode}
\usepackage{amsfonts}
\usepackage{amsmath}
\usepackage{amssymb}
\usepackage{booktabs}
\usepackage{caption}
\usepackage{cleveref}
\usepackage{enumitem}
\usepackage{float}
\usepackage{graphicx}
\usepackage{hyperref}
\usepackage{makecell}
\usepackage{multirow}
\usepackage{natbib}
\usepackage{subcaption}
\usepackage[table]{xcolor}
\usepackage{wrapfig}
\usepackage{xcolor}
\usepackage{pifont}
\usepackage{times}
\usepackage[many]{tcolorbox}  
\usepackage{latexsym}
\definecolor{cvprblue}{rgb}{0.21,0.49,0.74}

\usepackage{tikz}
\usepackage{xcolor}
\usepackage{newunicodechar}
\usepackage{pifont}
\newunicodechar{−}{\ensuremath{-}}

\title{LLMDistill4Ads: Using Cross-Encoders to Distill from LLM Signals for Advertiser Keyphrase Recommendations at eBay}

\author{Soumik Dey$^1$, Benjamin Braun$^1$, Naveen Ravipati$^1$, Hansi Wu$^1$, Binbin Li$^1$ \\
$^1$ eBay Inc\\}

\usepackage{multirow}
\usepackage{makecell}
\begin{document}
\maketitle
\begin{abstract}
E-commerce sellers are advised to bid on keyphrases to boost their advertising campaigns. These keyphrases must be relevant to prevent irrelevant items from cluttering Search systems and to maintain positive seller perception. It is vital that keyphrase suggestions align with seller, Search, and buyer judgments. Given the challenges in collecting negative feedback in these systems, LLMs have been used as a scalable proxy for human judgments. We present an empirical study on a major e-commerce platform of a distillation framework involving an LLM teacher, a cross-encoder assistant and a bi-encoder Embedding Based Retrieval (EBR) student model,  aimed at mitigating click-induced biases and provide more diverse keyphrase recommendations while aligning advertising, search and buyer preferences.
\end{abstract}

\maketitle% Ensure proper spacing adjustments are handled without misusing \vspace

\begin{comment}
\section{Introduction}
\begin{figure}[t]
\centering

\begin{subfigure}{8.5cm}
\centering
\includegraphics[width=\linewidth]{keyphrase.png}
  \caption{Buyer side}
  \label{fig:1a}
\end{subfigure}\qquad
\begin{subfigure}{8.5cm}
\centering

\includegraphics[width=\linewidth]{Sellerside.png}
\caption{Seller Side}
\label{fig:1b}
\end{subfigure}
\caption{Screenshot of our keyphrases for manual targeting in Promoted Listings Priority for  Advertising.}
\vspace{2mm}
\label{fig:screenshot}
\end{figure}
\end{comment}

\section{Introduction}

\begin{figure*}[t]
\centering
\includegraphics[width=\linewidth]{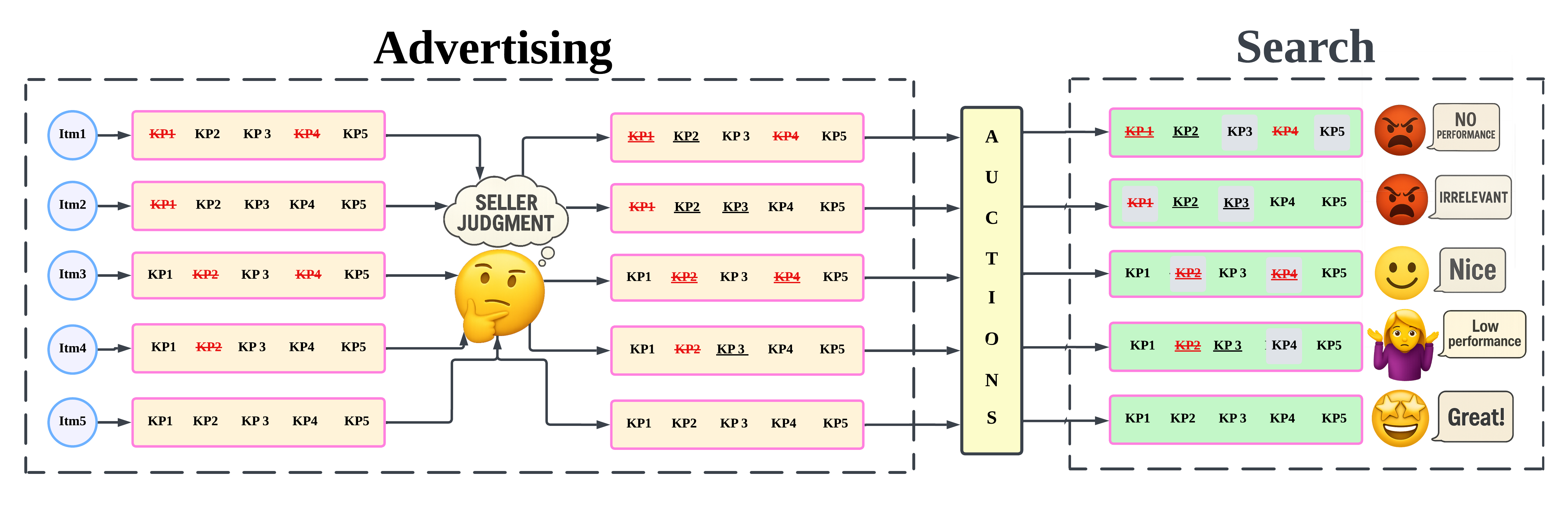}
\vspace{-8mm}
\caption{Auction mechanism of items (Itm) in relation to keyphrases (KP). Red strikethrough font represents filter of Advertising, the underline represents seller curation of keyphrases after advertising has filtered them while gray highlight represents the relevance filter of Search. }
\label{fig:auctions}
\vspace{-4mm}
\end{figure*}

In e-commerce, sellers rely on keyphrase recommendation systems \cite{mishra2025graphexgraphbasedextractionmethod, ashirbad-etal-2024, PUSL, mishra2025broadgenframeworkgeneratingeffective} to improve poor organic rankings and gain visibility on search results pages (SRPs). Because advertiser keyphrases correspond to buyer queries, their relevance is critical for effective seller strategies and for avoiding auction saturation with irrelevant items. Although high clicks or sales signal relevance, the absence of such signals does not imply irrelevance. E-commerce interaction data is typically missing-not-at-random (MNAR) \cite{surveybias, sampleselectionbias, clicksNquery, beyondPosBias, learning2rank, debiasedness, MNAR, lim2015top}, as buyer behavior is influenced by biased rankings. Items ranked lower—often less popular—receive fewer impressions, clicks, or purchases, making clicks and sales unreliable as negative relevance signals.

In sponsored search advertising, exact-match keyphrase recommendation systems suggest buyer queries for listings, enabling sellers to bid for sponsored slots on SRPs. User interactions on these pages are logged and used to infer query–item relationships, leading most approaches to frame keyphrase recommendation as an Extreme Multi-Label Classification (XMC) problem, with buyer queries as labels and click logs as ground truth \cite{dahiya2023ngame, renee_2023, Dahiya23b, you2019attentionxml, dahiya2021deepxml, ashirbad-etal-2024}. However, this formulation is plagued by a lot of challenges as described in the following paragraphs.

\paragraph{Click Data Biases and Evaluation} Click data is sparse and biased: 95\% of items receive no clicks, and clicks are affected by exposure, popularity, and selection biases~\cite{surveybias,sampleselectionbias}. The absence of clicks does not imply irrelevance~\cite{MNAR}. In addition, 90\% of the clicked items are associated with only a query, while the sellers expect 20–30 recommendations per item. Training on such data risks perpetuating bias and limiting coverage. Offline evaluation using precision/recall-based metrics reinforces existing biases and limits diversity. Sparse ground truths prevent meaningful evaluation beyond known associations, and XMC models (often stacked to produce more keyphrases) converge toward similar label predictions. As a result, improvements in offline metrics often do not translate into incremental performance \cite{mishra2025graphexgraphbasedextractionmethod}.

\paragraph{Keyphrase Targeting} Out-of-vocabulary (OOV) approaches for keyphrase generation using open-vocabulary models, such as GROOV \cite{simig-etal-2022-open}, One2Seq \cite{one2seq,one2set}, and One2One \cite{one2one1,chen-etal-2019-integrated,Chen_Gao_Zhang_King_Lyu_2019}, often propose keyphrases outside the predefined label set \cite{PUSL}. Generative models like TRIDENT \cite{tident} rely on an additional downstream Search matching algorithm to match between keyphrase and query. In contrast, closed or semi-open XMC models, including graph-based approaches \cite{mishra2025graphexgraphbasedextractionmethod, ashirbad-etal-2024} and embedding-based retriever systems \cite{DPR}, are preferable due to their inherently precise targeting, but have to be updated regularly to accommodate query drift, seasonality, and emerging trends.

\paragraph{Execution Performance} Real-time or daily batch inference requires low-latency, low-cost models. Heavy LLM-based approaches are at a disadvantage due to cost, latency, and scaling challenges due to regular inference updates and heavy inventory churn \textit{(inventory of 2.3 billion items with a daily churn of 20 million)} ~\cite{kaddour2023,MLSYS2023_LLMscaling,xin-etal-2020-deebert}. 

\paragraph{Relevance from different perspectives}
In e-commerce advertising, sellers bid on keyphrases via \textit{Advertising}, which are matched to buyer queries and filtered for relevance by \textit{Search} before entering auctions. Consequently, click logs contain only Search-approved keyphrases, creating a \textit{middleman bias}: models trained on click data never observe keyphrases rejected by Search, even though they are proposed by Advertising. This induces sample-selection bias~\cite{rec4ad,sampleselectionbias} and limits the reliability of click-based training~\cite{dey2025middlemanbiasadvertisingaligning}.

Relevance judgments also differ across sellers, Advertising, and Search. Sellers select keyphrases from Advertising, but Search may still reject them, causing items to miss auctions. Furthermore, even if Search or Advertising might find a keyphrase relevant but Sellers might reject them if they don't think it is relevant.  Figure~\ref{fig:auctions} depicts this with Itm1: while KP1 and KP4 are deemed irrelevant by Advertising, Search dismisses KP2, KP3, and KP5. The seller also discards KP2, causing Itm1 to miss auctions, regardless of the keyphrases’ relevance. Complete alignment is seen with Itm5, and partially with Itm3,
though some keyphrases still face exclusion. Ideally, Itm3 shouldn’t retrieve KP2 and KP4. Sellers ignoring KP2 for Itm1 can negate our suggestions, leading to irrelevant keyphrases, low satisfaction, and wasted resources.  Therefore, an effective keyphrase recommendation requires alignment between seller preferences, relevance of Advertising, and Search to improve adoption and campaign performance.

\section{Related Work}

Embedding-Based Retrieval (EBR) is a two-step process: buyer keyphrases and item titles are embedded into a shared vector space, and the k nearest keyphrases are retrieved for each item using approximate nearest neighbor (ANN) search. In semantic search, encoders are broadly categorized as bi-encoders and cross-encoders. Bi-encoders independently encode queries and items into fixed representations via self-attention, enabling efficient ANN retrieval. Cross-encoders, in contrast, jointly encode query–item pairs with cross-attention, capturing richer interactions at significantly higher computational cost.

Although both architectures are trained using supervised data, cross-encoders generally model complex query–document relationships more effectively. However, bi-encoders are preferred in EBR due to their ability to pre-compute embeddings for items and queries separately, making them scalable for large inventories. Fine-tuning bi-encoders requires labeled relevant and irrelevant query–item pairs. Training solely on click-based signals is problematic, as it reproduces exposure and popularity biases in the data \cite{surveybias}. Nonetheless, click-through-rate (CTR) logs provide reliable positive pairs \cite{huang2024piccolo2generaltextembedding}, while negative labels are less trustworthy as indicators of true irrelevance \cite{10.1145/2792838.2799671}.

Click data is further affected by middleman bias \cite{dey2025middlemanbiasadvertisingaligning}, a form of sample selection bias \cite{sampleselectionbias} introduced by bidding dynamics and Search–Advertising contracts. To mitigate these issues, some studies leverage Search Relevance (SR) signals to train relevance filters that validate keyphrase suggestions only when items are deemed relevant by Search. This approach has been shown to outperform click-based supervision \cite{dey2025middlemanbiasadvertisingaligning, dey2025judgejudgeusingllm}. Additionally, advertiser acceptance or rejection of suggested keyphrases highlights the need to align models with human judgment.

Large Language Models (LLMs) offer a scalable alternative for generating relevance labels. They can approximate human judgment while reducing bias and leveraging broad world knowledge, avoiding domain-specific tuning across large inventories \cite{liao2024d2llmdecomposeddistilledlarge}. Multi-task training further benefits from diverse supervision signals. For example, Piccolo2 \cite{huang2024piccolo2generaltextembedding} combines CoSENT loss \cite{Cosent} with a tailored InfoNCE loss that emphasizes hard negatives selected via BM25 \cite{10.1561/1500000019, DPR}.

Recent work explores LLMs as judges for large-scale label generation \cite{wang2024improvingpinterestsearchrelevance, gurjar2025dashclipleveragingmultimodalmodels, microsoft_llm_as_a_judge, gu2025surveyllmasajudge, DRAMA}. In particular, \cite{dey2025judgejudgeusingllm} shows that LLM-generated labels can effectively fine-tune cross-encoders for keyphrase relevance, outperforming models trained on search logs or click data. This line of work also emphasizes evaluating relevance models with business-oriented metrics to better reflect practical impact.

While cross-encoders achieve higher accuracy, their latency makes them unsuitable for large-scale retrieval. Consequently, substantial research focuses on distilling cross-encoder knowledge into more efficient bi-encoders. TwinBERT \cite{twinbert} and PROD \cite{prod} distill cross-encoders into twin-tower BERT architectures. ERNIE-Search \cite{lu2022erniesearchbridgingcrossencoderdualencoder} adopts a teacher–assistant framework \cite{mirzadeh2020improved}, transferring knowledge from a cross-encoder to a late-interaction model such as ColBERT \cite{10.1145/3397271.3401075}, and subsequently to a bi-encoder. CUPID \cite{Bhattacharya2023} observes that traditional pointwise MSE loss \cite{kim2021comparing} is ineffective for cross- to bi-encoder distillation. D2LLM \cite{liao2024d2llmdecomposeddistilledlarge} instead proposes multi-task distillation from an LLM-based cross-encoder, incorporating a Pearson correlation–based rank imitation loss as a more effective alternative to pointwise objectives.
 
\begin{figure*}[t]
    \centering
    \includegraphics[width=\textwidth]{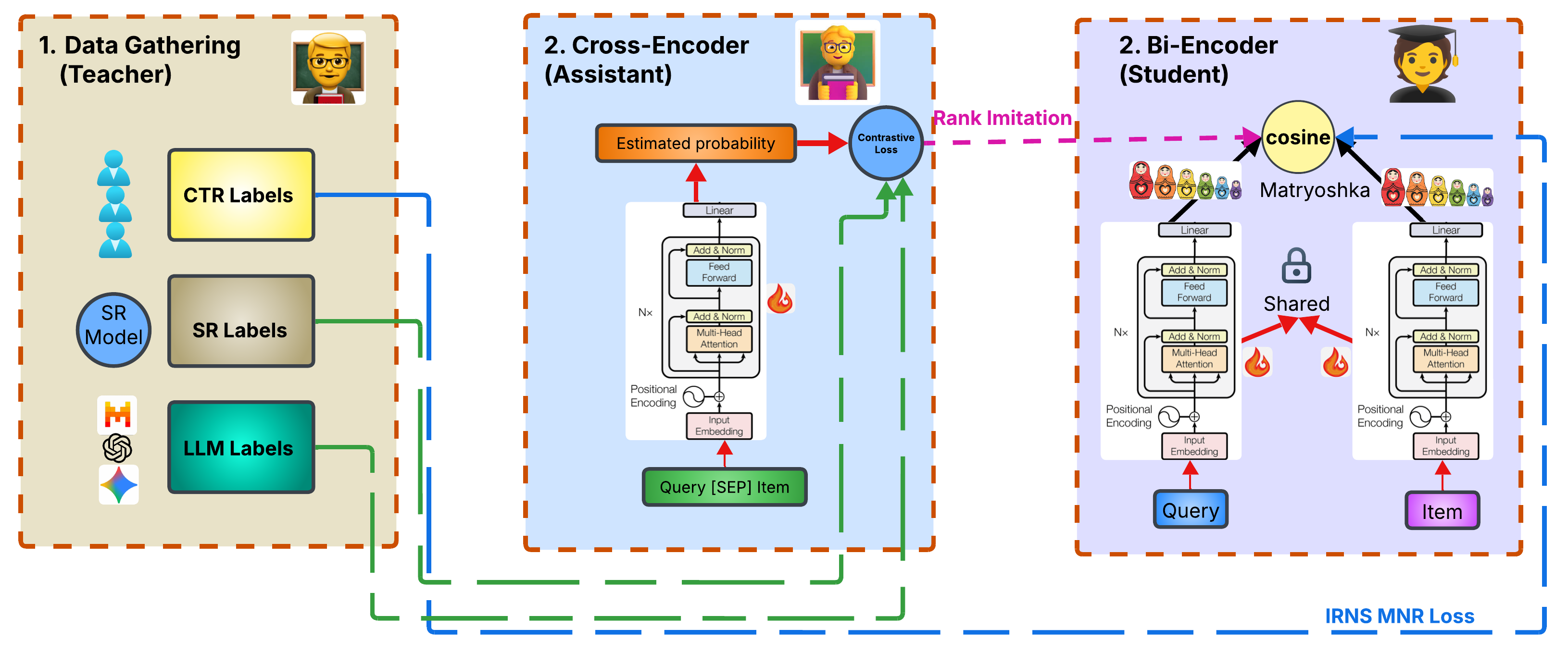}
    \caption{Our proposed architecture for multi-task knowledge distillation. The LLM is distilled to a cross-encoder, which is in turn distilled to the bi-encoder via multi-task hybrid training}
    \label{fig:architecture}
    \vspace{-3mm}
\end{figure*}
%Although D2LLM \cite{liao2024d2llmdecomposeddistilledlarge} makes no strong claims as to whether the Pearson-based rank imitation loss contributes to their performance (there were no ablation studies to dissect the performance), our hypothesis is that using this loss does contribute to the effective distillation from cross-encoder to bi-encoder over conventional MSELoss. 
\paragraph{Our contributions} We propose a framework to mitigate bias in click-derived data by training on heterogeneous supervision signals, including search relevance judgments and LLM-based labels. Our approach uses a multi-task learning framework augmented with a Teacher–Assistant hierarchy comprising an LLM teacher (approximating scalable human judgments), a cross-encoder assistant, and a bi-encoder student. We further conduct an empirical comparison of loss functions and distillation strategies for advertiser keyphrase retrieval, and introduce metrics and an offline evaluation setup that faithfully translates to  incremental online gains in stacked recall architectures.

\section{Embedding-Based Retrieval}
A dual-tower architecture that separately processes keyphrases and items offers an optimal foundation for a retrieval model aimed at cost-effective recommendations while maintaining reasonable latency. To address biases in click-based data, we enhance our dataset with auxiliary signals from Large Language Models (LLMs) on our own keyphrase recommendations and search relevance scores, adopting a hybrid training strategy as shown in Figure \ref{fig:architecture}. A cross-encoder assistant is further employed to distill from LLM-generated labels. We begin by describing our cross-encoder structure, then proceed to the bi-encoder design. Following this, we describe the multi-task training process employed to train/distill our bi-encoder for keyphrase retrieval.

\subsection{Dataset curation}

We compile data on user-query interactions via click-data, Search Relevance (SR) metrics, and relevance scores from Large Language Models (LLM), all based on item-keyphrase pairs at eBay. Click-data stems from item-queries that pass a search relevance filter (\textit{middleman bias}) and are further affected by ranking biases. In contrast, SR and LLM datasets of item-keyphrase pairs are devoid of sample selection or intermediary biases since they are collected from upstream signals to clicks. \footnote{We also track whether sellers adopt our keyphrases, but this signal is noisy: rejection doesn’t clearly reflect a dislike of the phrases themselves, since they come bundled with a bid. When sellers don’t adopt, we can’t tell if they’re rejecting the bid, the keyphrases, or both.} 

\paragraph{\textbf{CTR-based labels}}

For each (query, item) pair, we compute the CTR as the ratio of clicks to impressions over the past 30 days of eBay's search logs. A pair is labeled positive when this ratio exceeds 0.05. To mitigate label noise (1 click 1 impression or 1000 impressions 1 click) we enforce minimum thresholds on CTR, clicks, and impressions. The resulting click-labeled dataset comprises 10,702,747 instances.

\paragraph{\textbf{Search Relevance labels}}

We collected relevance scores during the auction for item-keyphrase pairs for 3 months. These are generated by our Search Relevance (SR) model, and are averaged for each (item, keyphrase) pair. A (query, item) pair is labeled positive if its SR score exceeds a threshold set by business metrics, varying by country, and is negative otherwise. This training dataset contains 18,721,682 records. 

\paragraph{LLM labels}
We obtained relevance labels for each (item, keyphrase) pair using Mixtral 8x7B Instruct-v0.1 \cite{jiang2024mixtralexperts}. This model shows 90\% agreement with click data—used here as a proxy for positive human feedback—and exhibits reasonable consistency with separately collected human judgments (see Appendix \ref{sec: bench}). The training set is the same as the SR-set with a test set of 3,524,414 instances.

\begin{comment}
\begin{tcolorbox}[colback=gray!5!white, colframe=black, title=Prompt Design]
\noindent\texttt{\footnotesize{Below is an instruction that describes a task. Write a response that appropriately completes the request. \\
\\
\#\#\# Instruction: \\
Given an item with title: "\{\textrm{title}\}", determine whether the keyphrase: "\{\textrm{keyphrase}\}", is relevant for cpc targeting or not by giving ONLY yes or no answer: \\
\\
\#\#\# Response:
}}
\end{tcolorbox}
\end{comment}

\subsection{The cross-encoder}
For the cross-encoder (CE) inputs, one input is the user-provided keyphrase, while the second consists of both the item title and its corresponding category, combined together. Consequently, the comprehensive input delivered to the cross-encoder is structured as \texttt{query [SEP] category name [SEP] item title}. 

The base model we used for the cross-encoder is the microBERT model, a distilled version of eBERT (pre-trained on a dataset that includes item title corpus from eBay) with the architecture of mobileBERT \cite{DBLP:journals/corr/abs-2004-02984}. It is a compact and efficient version of eBERT that retains high accuracy while significantly reducing model size and inference latency. More precisely, it is 4.3× smaller and 5.5× faster than eBERT while achieving comparable performance.
We fine-tuned the cross-encoder on the labels coming from the LLM model of 50,078,315 records, with the cross-entropy loss on the dataset described above. When evaluated on a test set of 7,503,031 (item, keyphrase) pairs, it achieved a F1 score of 96\%, thus validating its use as an assistant model. 

\subsection{The bi-encoder}
The bi-encoder (BE) model which also uses microBERT as base, encodes item titles with their meta category (\texttt{item title [SEP] category name}) and advertiser keyphrases (buyer query) separately. These are compared using \textit{cosine similarity} following a mean pooling operation. For ANN latency purposes the embeddings were further truncated using Matryoshka Loss \cite{kusupati2024matryoshkarepresentationlearning} to 64. %The choice of the base model significantly influences the quality of the derived embeddings. In the next section, we present the different base models that we have tried as part of our experiments.

\section{Results}

In our experimental design, we sought to refine a student bi-encoder model employing a training methodology that melds rank imitation loss derived from the output of an assistant cross-encoder with a multi-task training strategy influenced by distinct ground truth labels. We utilized separate loss functions tailored to each label and integrated them in a multi-task framework where each batch solely includes samples from one dataset. Training and evaluation datasets were sampled proportionally to their size, influenced by the framework introduced in Picolo2 \cite{huang2024piccolo2generaltextembedding}. We consider a diverse set of training objectives that span Multiple Negatives Ranking Loss (MNR) \cite{MNR} using In Batch Random Negative Sampling (IRNS) for CTR labels; Contrastive Loss \cite{contrastive} and Softmax Loss \cite{reimers2019sentencebertsentenceembeddingsusing} for $LLM \rightarrow BE$ labels distillation; Mean Squared Error (MSE: \citealt{kim2021comparing}), CoSENT \cite{Cosent}, Margin MSE ($\mathrm{MSE}_{mar}$: \citealt{hofstätter2021improvingefficientneuralranking}), Pearson Correlation Loss \cite{liao2024d2llmdecomposeddistilledlarge}, and KL-Divergence Loss for $LLM \rightarrow CE  \rightarrow BE$ score distillation (more details in the Appendix \ref{app:loss}).

\definecolor{green1}{RGB}{30,70,30}     % Lowest value
\definecolor{green2}{RGB}{80,130,80}
\definecolor{green3}{RGB}{120,170,120} 
\definecolor{green4}{RGB}{200,230,200}
\definecolor{green5}{RGB}{235,250,235}  % Highest value

\subsection{Ablation on Knowledge Distillation Losses}

We first analyze the knowledge distillation (KD) mechanism, which forms the backbone of our framework. In our setting, a bi-encoder (BE) is trained to mimic the soft predictions of a cross-encoder (CE), which itself is distilled from an LLM. The objective of KD is therefore not merely binary classification agreement, but faithful transfer of ranking behavior and score calibration from teacher to student.

To assess alignment between the BE and CE, we report F1, precision (P), recall (R), and the Pearson correlation coefficient ($\rho$). While F1, P, and R evaluate classification quality, Pearson correlation measures agreement between the BE and CE soft scores, directly capturing calibration and ranking alignment under knowledge distillation.

Table~\ref{tab:results2} compares different KD objectives under both two-stage distillation $\mathrm{LLM} \rightarrow \mathrm{CE} \rightarrow \mathrm{BE}$ and direct distillation 
$\mathrm{LLM} \rightarrow \mathrm{BE}$. We observe that the two-stage pipeline consistently outperforms direct distillation. For example, under Pearson loss, the two-stage setup achieves F1 = 0.88 and $\rho = 0.87$, whereas direct contrastive distillation yields F1 = 0.83 and $\rho = 0.76$, and softmax distillation degrades further (F1 = 0.66, $\rho = 0.45$). This demonstrates that the CE acts as an effective calibration intermediary when transferring LLM knowledge.

Among KD losses, mean squared error (MSE) performs worst (F1 = 0.81, $\rho = 0.78$), corroborating prior findings that regression-style objectives are suboptimal for ranking transfer. CoSENT, a pairwise ranking objective with calibration, improves performance (F1 = 0.87, $\rho = 0.82$). The Pearson correlation loss achieves the best overall results (F1 = 0.88, P = 0.87, R = 0.88, $\rho = 0.87$), indicating that directly optimizing for correlation better preserves both ranking order and score calibration. KL-divergence and margin-based variants provide intermediate performance but do not surpass Pearson. \textit{These findings suggest that batch-wise ranking imitation aligned with correlation objectives is most effective for KD.} We therefore adopt the Pearson-based loss in subsequent experiments.
\begin{table}[t]
\footnotesize
\setlength{\tabcolsep}{3pt} % tighten column spacing
\centering
\begin{tabularx}{\linewidth}{@{}l|l||*{3}{>{\centering\arraybackslash}X}>{\centering\arraybackslash}X@{}}
\toprule
\rowcolor{gray!20}
\textbf{KD} & \textbf{Loss} & \textbf{F1} & \textbf{P} & \textbf{R} & \(\boldsymbol{\rho}\) \\
\midrule
\multirow{5}{*}{\makecell{$\mathrm{LLM} \rightarrow \mathrm{CE}$\\ $\rightarrow \mathrm{BE}$}}
& $\mathrm{MSE}$       & \cellcolor{green5}{0.81} & \cellcolor{green3}{0.77} & \cellcolor{green3}{0.86} & \cellcolor{green3}{0.78} \\ %0.63
& CoSENT    & \cellcolor{green3}{0.87} & \cellcolor{green3}{0.86} & \cellcolor{green1}{\textcolor{white}{0.88}} & \cellcolor{green2}{\textcolor{white}{0.82}} \\ 
& Pearson   & \cellcolor{green1}{\textcolor{white}{0.88}} & \cellcolor{green1}{\textcolor{white}{0.87}} & \cellcolor{green1}{\textcolor{white}{0.88}} & \cellcolor{green1}{\textcolor{white}{0.87}} \\ % spear=0.87
& $\mathrm{MSE}_{mar}$ & \cellcolor{green4}{0.86} & \cellcolor{green4}{0.84} & \cellcolor{green1}{\textcolor{white}{0.88}} & \cellcolor{green2}{\textcolor{white}{0.80}} \\ %0.80
& KL-Div    & \cellcolor{green4}{0.85} & \cellcolor{green4}{0.83} & \cellcolor{green1}{\textcolor{white}{0.88}} & \cellcolor{green4}{0.66} \\ %sp- 67
\midrule
\multirow{2}{*}{\makecell{$\mathrm{LLM} \rightarrow \mathrm{BE}$}}
& Contrastive    & \cellcolor{green3}{0.83} & \cellcolor{green3}{0.80} & \cellcolor{green3}{0.87} & \cellcolor{green3}{0.76} \\ %0.81
& Softmax   & \cellcolor{green4}{0.66} & \cellcolor{green3}{0.60} & \cellcolor{green5}{0.73} & \cellcolor{green5}{0.45} \\ %spear - 0.53
\bottomrule
\end{tabularx}
\caption{Ablation on changing the KD Loss}
\label{tab:results2}
\vspace{-3mm}
\end{table}

\subsection{Label Ablation and Offline Evaluation}

Having identified the optimal KD configuration, we next examine how different supervision signals contribute under production-like settings (Table~\ref{tab:results3}) from a sampled subset of items (1000). We assess models using the following metrics.

\paragraph{Median Keyphrases per Item (KP)}
In deployed systems, recommendations typically originate from several layered recall sources. Whereas most of these sources are trained on click-based signals, our framework distills knowledge from an LLM, introducing an indirect debiasing mechanism. To assess the contribution of our recall source, we count only those keyphrases it generates that are not already produced by other active recall sources, and that pass the downstream Advertising Relevance Filter. We then compute the median number of unique, de-duplicated, relevant keyphrases per item. \textit{KP quantifies incremental utility and efficiency, i.e., the additional high-quality suggestions supplied by the model.}

\paragraph{LLM Pass Rate (PR)}
Over the surfaced keyphrases (see paragraph above), we measure the percentage that are accepted by our LLM-based judge (Mixtral-8x7B-Instruct-v0.1), which is used as a proxy for seller evaluation. \textit{PR quantifies alignment of surfaced keyphrases with LLM judgment which serves as a proxy for human sentiment.}

\paragraph{Cumulative Retrieval Pass Rates}
Beyond evaluating the deduplicated setting, we also aimed to measure the standalone ranking and recall performance of our framework. Using top-20 retrieved keyphrases (non-deduped), we compute the cumulative proportion that satisfy LLM judgment at cutoffs \@ 5, 10, 15, and 20 per-item. In the absence of graded relevance ground truth, these statistics approximate ranking quality at varying depths and reveal how well the system concentrates relevant keyphrases near the top of the ranked output.

\textbf{CTR-only} yields KP = 7 and PR = 60. Incorporating \textbf{only LLM} supervision raises KP to 11 but leaves top-rank precision largely unchanged. \textbf{KD-only} performs poorly (PR = 39), indicating that distillation without additional supervision is inadequate. Adding auxiliary signals systematically enhances performance: \textbf{CTR+LLM} increases PR to 69 but lowers KP to 6, showing adding only CTR to LLM lowers diversity. \textbf{LLM+KD} improves upon \textbf{KD-only} (PR = 49), showing that distillation is beneficial when combined with semantic guidance. \textbf{LLM+SR+KD} further improves both coverage and ranking concentration (KP = 12, PR = 51). Most importantly, \textit{the addition of LLMs sources improves effective coverage (KP)} (except CTR+LLM), signifying the effect of LLMs in debiasing and providing more diverse recommendations.

The best outcomes arise when LLM, CTR, and KD are used jointly. \textbf{LLM+CTR+KD} attains KP = 12 (highest median incremental keyphrases), PR = 71 (strongest seller-proxy approval), and top-rank pass rates of 68\% @5, 60\% @10, 55\% @15, and 52\% @20. \textbf{LLM+SR+CTR+KD} is comparable (PR = 70, KP = 11) but slightly worse at shallow depths. Gains are most pronounced at the top of the ranking: improving from \textbf{CTR} (51\% @5) to \textbf{LLM+CTR+KD} (68\% @5) yielding a marked increase in ranking concentration. All variants retain $>$99\% agreement with the Search Relevance model, confirming that higher LLM approval does not degrade auction efficiency. \textit{In summary, Pearson-based KD, when combined with LLM and CTR supervision, offers the best trade-off among calibration, ranking quality, and incremental utility and diversity}.

\begin{table}[t]
\centering
\footnotesize
\begin{tabular}{l||cccccc}
\toprule
\rowcolor{gray!20}
\textbf{Model} & \textbf{KP} & \textbf{PR} & \textbf{5} & \textbf{10} & \textbf{15} & \textbf{20} \\
\midrule

CTR               & \cellcolor{green4}{7} & \cellcolor{green3}{60}
                  & \cellcolor{green3}{51} & \cellcolor{green3}{42}
                  & \cellcolor{green3}{37} & \cellcolor{green4}{34} \\

LLM               & \cellcolor{green2}{\textcolor{white}{11}} & \cellcolor{green2}{\textcolor{white}{61}}
                  & \cellcolor{green3}{45} & \cellcolor{green4}{41}
                  & \cellcolor{green3}{38} & \cellcolor{green3}{35} \\
KD                & \cellcolor{green2}{\textcolor{white}{11}} & \cellcolor{green5}{39}
                  & \cellcolor{green5}{29} & \cellcolor{green5}{27}
                  & \cellcolor{green5}{26} & \cellcolor{green5}{25} \\

CTR+LLM           & \cellcolor{green5}{6} & \cellcolor{green2}{\textcolor{white}{69}}
                  & \cellcolor{green3}{{57}} & \cellcolor{green3}{{48}}
                  & \cellcolor{green3}{{43}} & \cellcolor{green3}{{39}} \\

SR+LLM             & \cellcolor{green2}{\textcolor{white}{11}} & \cellcolor{green4}{46}
                  & \cellcolor{green4}{36} & \cellcolor{green5}{34}
                  & \cellcolor{green5}{32} & \cellcolor{green5}{31} \\

LLM+KD            & \cellcolor{green2}{\textcolor{white}{11}} & \cellcolor{green4}{49}
                  & \cellcolor{green4}{36} & \cellcolor{green4}{35}
                  & \cellcolor{green4}{33} & \cellcolor{green5}{32} \\

LLM+SR+KD         & \cellcolor{green1}{\textcolor{white}{12}} & \cellcolor{green3}{51}
                  & \cellcolor{green3}{47} & \cellcolor{green3}{42}
                  & \cellcolor{green3}{41} & \cellcolor{green3}{39} \\

LLM+CTR+KD        & \cellcolor{green1}{\textcolor{white}{12}} & \cellcolor{green1}{\textcolor{white}{71}}
                  & \cellcolor{green1}{\textcolor{white}{68}} & \cellcolor{green1}{\textcolor{white}{60}}
                  & \cellcolor{green1}{\textcolor{white}{55}} & \cellcolor{green1}{\textcolor{white}{52}} \\

LLM+SR+CTR+KD     & \cellcolor{green2}{\textcolor{white}{11}} & \cellcolor{green2}{\textcolor{white}{70}}
                  & \cellcolor{green2}{\textcolor{white}{67}} & \cellcolor{green2}{\textcolor{white}{59}}
                  & \cellcolor{green2}{\textcolor{white}{54}} & \cellcolor{green2}{\textcolor{white}{51}} \\

\bottomrule
\end{tabular}
\caption{Combined comparison of label ablation metrics (KP, PR) and cumulative retrieval pass rates till ranks 5, 10, 15, and 20.}
\vspace{-3mm}
\label{tab:results3}
\end{table}

\setlength{\textfloatsep}{15pt} 

\begin{comment}
\section{Production System Design}
The production architecture depicted in Figure \ref{fig:prod_arch} comprises two main parts: \textit{Near Real-Time (NRT)} Inference and Batch Inference. Batch inference handles items with a delay, while NRT prioritizes immediate items, particularly those newly created or updated by sellers. Batch inference has two components: 1) full batch inference for all items, and 2) daily differential (Diff) to integrate new and updated items with existing data. NRT inference utilizes triton and onnx serving using V100 GPUs, activated by item creation or updates managed by Flink processing and feature enrichment. The full batch handles approximately 2.3 billion items, while the daily Diff supports a churn of 20 million items. As the full batch runs just once, Diff latency determines model deployment viability, being about 35 minutes for bi-encoders. The ANN job downstream on a takes an additional 2.5 hours daily and for NRT our vector database service helps in that regard. Latency numbers reported for our batch inference use PySpark \cite{spark} (1500 executors, 20g memory, 4 cores), leveraging transformers \cite{wolf-etal-2020-transformers} and onnxruntime \cite{onnxruntime}.
\end{comment}
\subsection{Online Impact}
Over a 12-day A/B test in the US market, replacing the \textbf{CTR-only} EBR model with \textbf{LLM+CTR+KD} produced substantial improvements: \textit{gross merchandise volume bought} (GMB; seller sales) increased by $51.26\%$ ($p=0.01$), \textit{return on advertising spend} (ROAS; GMB-to-ad-spend ratio) increased by $38.69\%$ ($p=0.02$), and the average adopted keyphrase count per item rose by $11.75\%$ ($p=0.03$), indicating higher seller adoption.

\section{Conclusion}
This study investigates the limitations of solely using click-based cues to refine bi-encoder models for classifying sentence pairs in e-commerce. We found that supplementing with signals from LLMs significantly boosts model performance and diversity in stacked model architectures conventionally reliant only on click-based training data. Further improvements are achieved by incorporating an intermediate cross-encoder model using knowledge distillation during fine-tuning. Notably, combining this distillation with training on CTR, and raw LLM labels enhances bi-encoder efficacy. We also observe that, Pearson correlation Loss --- a rank imitation loss based on Pearson correlation is a superior knowledge distillation loss function, surpassing a plethora of other losses in our use case. Lastly, we offer an evaluation protocol designed to measure the business impact of retrieval models in a two-sided marketplace setting of advertising, emulating real-world production settings.

%\section{Limitations}
%This study does an in depth analysis of using LLM labels for mitigating click based biases present in Advertisement systems which typically involve a two-sided marketplace. While this is an empirical study involving huge customer base and an inventory of billions of items, the study is quite niche and limited to the Advertisement space. In addition, the use of general purpose LLMs is necessitated by the defects in procurement of quality human judgment data (see Appendix \ref{sec: bench}), affected by modality biases. In the future with the procurement of better human judgment data could steer this in a direction involving fine-tuned LLMs. In the CTR dataset, more complex negative mining techniques could have been explored, like ANCE~\cite{xiong2021approximate} and N-Game~\cite{dahiya2023ngame}, however due to the size of the dataset and brevity of our study, we leave the exploration of cost-effective negative mining strategies to reinforce our signals from interaction data to future research. 

%\end{document}
% \bibliographystyle{plainurl}
\bibliography{main}
\section{Appendix}
\subsection{Benchmarking LLM-as-a-Judge}
\label{sec: bench}

There are predominantly two principal strategies for utilizing LLMs as judges in data generation and augmentation:

\begin{itemize}
    \item \textbf{General LLM:} Advanced models such as GPT-4 \cite{NEURIPS2022_b1efde53, openai2024gpt4technicalreport} are viable alternatives to human judgment \cite{NEURIPS2023_91f18a12} in LLM-as-a-Judge automation. \cite{alpaca_eval} developed an 805-question benchmark to compare performance with text-davinci-003 using GPT-4\cite{NEURIPS2023_91f18a12}. \cite{zhu2025judgelmfinetunedlargelanguage} created 80 multi-round queries over eight fields, using GPT-4 for automated scoring. Despite the accuracy and reliability of closed-source models like GPT-4, Gemini 2.5 \cite{comanici2025gemini25pushingfrontier}, Claude 3 \cite{anthropic2024claude3} etc, surpassing human evaluations, its usage is often restricted by rate limits or API access. The Mixtral-7x8B Instruct v0.1 \cite{jiang2024mixtralexperts} model effectively annotates keyphrase relevance in advertising. Its open-source nature facilitates model distillation and training and its medium size is essential to produce substantial amounts of judgments needed for covering the diversity of our platform's inventory without too much GPU acquistion costs. \footnote{LLAMA 2 \cite{touvron2023llama2openfoundation}, DBRX \cite{databricksIntroducingDBRX}, and Qwen-2 \cite{yang2024qwen2technicalreport} were considered during development but faced distillation, legal and licensing challenges for commercial deployment.} While a general LLM's shortcomings in instruction adherence or reasoning may undermine its utility as a judge, its extensive knowledge base helps avoid biases inherent in fine-tuned models \cite{lichtenberg2024largelanguagemodelsrecommender}.
    \item \textbf{Fine-tuned LLM:} Fine-tuning a judge model involves several steps \cite{huang2024empiricalstudyllmasajudgellm}: (1) Data Collection, which includes gathering instructions, subjects, and evaluations, typically using data sets and annotations from GPT-4 or humans. (2) Prompt Design, adapting templates for evaluation. (3) Model Fine-Tuning, using prompts and data to instruct the model via an instruction tuning framework. Post-tuning, the model can evaluate target entities. Despite enhanced test set outcomes, these models have evaluation constraints and retain biases from human annotators \cite{gu2025surveyllmasajudge}. Issues such as improper prompt and dataset design can impair generalization, complicating comparisons with robust models like GPT-4.
\end{itemize}

\begin{figure*}[t]
\includegraphics[width=\linewidth]{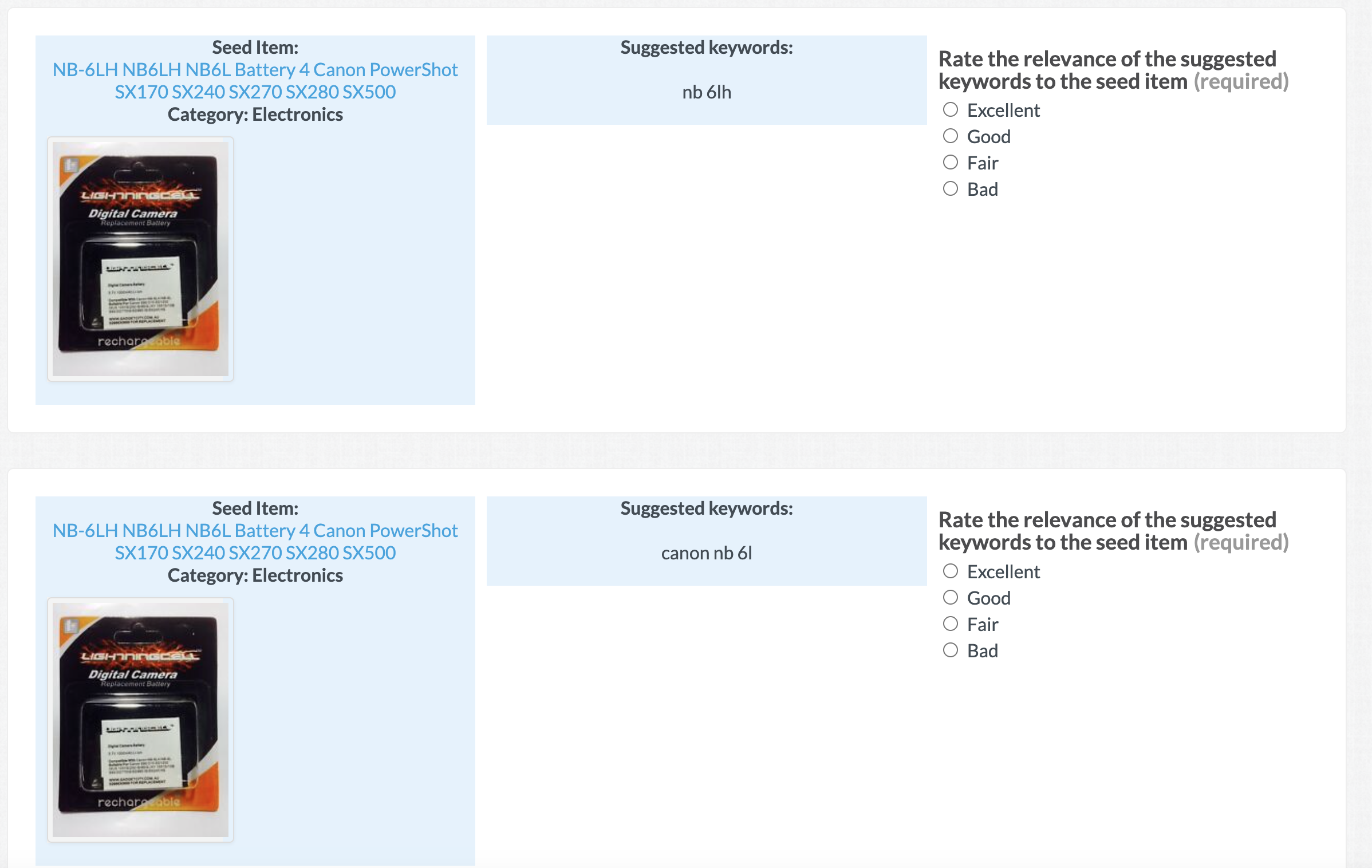}
\caption{Interface for our human annotators.}
\label{human}
\end{figure*}

In order to obtain a set of scores from a large language model (LLM) as either an augmentation or a substitute for human evaluation, we collected around 50 million judgment scores derived from the Mixtral 8x7B Instruct v0.1 model. This process utilized two distinct versions of the Mixtral 8x7B Instruct v0.1: 1) the standard general-purpose LLM, and 2) a version fine-tuned specifically to binary human judgment labels. The prompt is illustrated below.

\begin{tcolorbox}[colback=gray!5!white, colframe=black, title=Prompt Design]
\noindent\texttt{\footnotesize{Below is an instruction that describes a task. Write a response that appropriately completes the request. \\
\\
\#\#\# Instruction: \\
Given an item with title: "\{\textrm{title}\}", determine whether the keyphrase: "\{\textrm{keyphrase}\}", is relevant for cpc targeting or not by giving ONLY yes or no answer: \\
\\
\#\#\# Response:
}}
\end{tcolorbox}

When evaluated against 150,000 human judgment scores (this was human judgment that was asked by human annotators to label our recommendations as shown in Figure \ref{human} and instructions below) collected in the course of our analysis, the general LLM shows a fair level of agreement, with a kappa coefficient of 0.258. In contrast, the fine-tuned LLM exhibits superior alignment with human-provided judgments, achieving a kappa coefficient of 0.724.
\begin{tcolorbox}[colback=gray!5!white, colframe=black, title=Ads Keyword Suggestion for Evaluation]

\scriptsize{
Sellers participating in the Promoted Listings ads program can bid on search query keywords and phrases where they want their items to appear. Help us to evaluate the performance of an automated tool that suggests keyword phrases for which a seller might want their ad to appear. How relevant are the suggested keywords to the item?

\textbf{Question: How relevant are the suggested keywords to the seed item?}

\begin{itemize}[leftmargin=*]
    \item \textbf{Excellent} -- the suggested keywords are highly relevant and match the seed item in all core attributes.
    \item \textbf{Good} -- the suggested keywords are somewhat relevant and provide an OK or good enough match to the item. Some of the secondary attributes might not match, but important traits like model, size, or product type are respected.
    \begin{itemize}
        \item Note, for the purposes of ad placement, keywords containing a competing brand and similar model can qualify as Good.
        \item Example: seed item is a Samsung Galaxy phone, suggested KWs are ``google pixel pro.'' Both \textbf{phones} run on Android and are flagship models so have enough similarity to qualify as Good.
    \end{itemize}
    \item \textbf{Fair} -- the suggested keywords are only slightly relevant and miss the target in an important way. Core traits like product type or size might not match, but you can understand the connection between the seed item and suggested keywords.
    \item \textbf{Bad} -- the suggested keywords are not at all relevant and do not match the seed item on most, if not all, significant traits.

\end{itemize}
}
\end{tcolorbox}

For evaluating the LLM-as-a-judge framework, we employed click data as a dependable benchmark as described in other works such as \citealt {mishra2025graphexgraphbasedextractionmethod, dey2025judgejudgeusingllm} etc. Click data effectively signifies positive relevance, ensuring a robust evaluation. The LLM Mixtral 8x7B Instruct v0.1 used in this research aligned strongly with click-based human-generated positive annotations, surpassing 90\% agreement with click signals, indicating strong real-world performance. In this study since it involves a distillation process using cross-encoders, we assessed judgment effectiveness by calculating threshold values for cross-encoder for relevance judgments distilled from the general and fine-tuned LLMs, ensuring 95\% of clicks from search logs were retained~\footnote{The 95\% threshold is based on business logic and is implemented in current production models.}. Using these thresholds, we evaluated: 1) potential sales revenue loss with these thresholds applied to the same click data, 2) reduction in keyphrase volume when applied to our keyphrase recommendations, and 3) the \textit{search pass rate}, or the fraction of keyphrases passing both the Advertising and search relevance filters. An optimal model minimizes sales loss, maintains consistent click performance, and reduces recommended keyphrases while ensuring efficient search pass rates. The cross encoder distilled from fine-tuned LLM retained 75\% more keyphrases but reduced sales by 20\% with the same clicks, indicating suboptimal alignment with buyer judgment. In contrast, the cross-encoder distilled from the general LLM reduced 68\% more keyphrases, retained the same clicks, and increased sales by 10\%. The search pass rate is the same for both. These outcomes led us to choose the cross-encoder distilled from the general LLM model over its counterpart distilled from fine-tuned LLM, due to its strong alignment with buyer judgment across an extensive amount of data and ignoring the small sample of human judgment data we collected.

The practice of implementing human judgment frameworks across various projects in the industry is common. However, our experiments indicate potential pitfalls in this approach. Diverging from the norm where fine-tuned LLMs act as data augmentors or evaluators, our findings prompted an analysis to uncover the cause. We identified issues with our human judgment data: \begin{itemize}
    \item Annotators rated keyphrase pairs with labels such as \textit{excellent}, \textit{good}, \textit{fair}, and \textit{bad}. Despite clear instructions, the open-ended nature and complexity of these labels likely hampered judgment. A binary assessment would align better with our aim of binary classification.
    \item Annotations were insufficient for the vast array of eBay's listings and diverse buyer queries encountered daily.
    \item Annotators viewed item images which the non-multimodal models could not access, leading to modality bias \cite{park2024assessingmodalitybiasvideo, 10.1145/3565266, dey24_speechprosody}. Both annotators and models should have uniform input modalities. For instance, an item titled \texttt{``iPhone 11 64GB 128G Unlocked ATT Boost Cricket Spectrum Excellent Condition''} is yellow in the image, influencing annotators to deem \texttt{``yellow iphone''} relevant but \texttt{``red iphone''} irrelevant. Without visual cues, such data mislead the models. In addition the downstream cross-encoder and bi-encoder models are also not multimodal (due to latency and other issues, like doing inference on 2.3 billion items with images) which also motivated our choice to keep the LLMs text-only.
\end{itemize}

Developing improved methods for obtaining human judgments is vital, yet creating unique designs for each individual use case is impractical due to the extensive web data and eBay's inventory of 2.3 billion items across diverse products. The high cost of human evaluations makes it unfeasible to customize designs for every scenario. Furthermore, when large language models (LLMs) are fine-tuned on such data, the dataset's inherent biases often amplify and propagate through subsequent models, causing inaccuracies. Thus, a thorough assessment of data usage and the need for a fine-tuned LLM-as-a-judge is critical. It may be more beneficial to use a general-purpose LLM with a rigorously developed evaluation framework based on business metrics.

Additionally, challenges arise in the distillation process and click thresholding among various models. For example, while both LLMs find the item \texttt{``Genuine 15V 4A Power AC Adapter Laptop Charger For Surface Pro 3 4 5 6''} and the keyphrase \texttt{``microsoft surface charger''} relevant, the distilled fine-tuned LLM disagrees, unlike the original LLM. The cross-encoder fine-tuned model shows reduced calibration accuracy, excluding high-conversion items for a 95\% click threshold. In contrast, the distilled model from its general counterpart demonstrate better calibration, maintaining clicks, and sales effectively.

\subsection{Choice of base model}

We have experimented with 3 models, eBERT, MicroBERT and ModernBERT.  

\textbf{eBERT}: Multilingual BERT model \cite{dahlmann2021deployingbertbasedquerytitlerelevance}, pre-trained on our platform's item data and general domain. The architecture used is a BERT-base configuration with 12 layers, which produces an embedding of dimension 768. 

\textbf{MicroBERT}: Compressed and distilled version of eBERT (around 4 times smaller, and around 5 times faster; trained with a procedure explained in \citealt{DBLP:journals/corr/abs-2004-02984}). It achieves a smaller size due to a smaller intermediate layer (size of the feedforward layer inside the transformer), of 384 compared to 3072 for the original; Output embedding dimension is still 768.

\textbf{ModernBERT}: \cite{warner2024smarterbetterfasterlonger}
We used a version of modernBERT that is made multilingual through trans-tokenization and cross-lingual vocabulary transfers \cite{remy2024transtokenizationcrosslingualvocabularytransfers}. However, this base model was \textit{not pretrained on the specific platform data}. ModernBERT features many improvements over the original BERT architecture, including: longer sequence length (8192 tokens, compared to 512 for original BERT), the use of rotary positional embeddings instead of absolute ones, alternating global and local attention (every third layer uses global attention; the rest use local sliding window attention - all of these use Flash attention). Generally speaking, it is deemed a better model than the original BERT model, with an overall GLUE score of 88.5 (compared to 80.5 for BERT-base).

In an ablation study for selecting our base model, we have fine-tuned our bi-encoder model using the LLM-labeled training set with the contrastive loss function. The only parameter that we have changed here is the base model. We report the classification metrics we observed when using microBERT, modernBERT and eBERT as our base model.

As shown in Table \ref{tab:results1}, both the microBERT and the eBERT models give better performance than modernBERT, even though modernBERT has a higher GLUE score than BERT (88.5 vs 80.5) and a much higher context length (8192 vs 512). This result illustrates \textit{the importance of pre-training}, as the modernBERT version that we used here was not pre-trained on our platform-specific vocabulary.

Table \ref{tab:results1} also shows that using eBERT as our base encoder yields slightly better results than microBERT. This is expected, as microBERT is a distilled version of eBERT. Due to the size of our dataset, we chose to use microBERT for the rest of this study, as it shortens the batch prediction time by 30\%.

\begin{table}[t]
\centering
\begin{tabular}{@{}c||c|c|c@{}}
\toprule
\rowcolor{gray!20}
\textbf{Base models} & \textbf{Recall}   & \textbf{Precision} & \textbf{F1}       \\ 
\midrule
MicroBERT   & \cellcolor{green1}{\textcolor{white}{0.92}} & \cellcolor{green3}{0.78}             & \cellcolor{green3}{0.85}            \\
eBERT       & \cellcolor{green1}{\textcolor{white}{0.92}} & \cellcolor{green1}{\textcolor{white}{0.81}} & \cellcolor{green1}{\textcolor{white}{0.86}} \\
ModernBERT  & \cellcolor{green3}{0.91}             & \cellcolor{green5}{0.76}             & \cellcolor{green5}{0.83}            \\
\bottomrule
\end{tabular}
\caption{Changing the base model}
\label{tab:results1}
\end{table}

\subsection{Loss Functions}
\label{app:loss}
\paragraph{Multiple Negatives Ranking Loss}

The Multiple Negatives Ranking (MNR) Loss \cite{MNR} is well-suited to cases where only positive pairs are available, as it does not require manually labeled negative samples. When fed with item-keyphrase pairs of positive examples, this loss uses one item as its anchor, uses its given keyphrase as a positive example, and considers all other keyphrases in the training batch as negative for this anchor item (IRNS). This approximation works well with highly-sparse datasets such as e-commerce and web datasets. In our use-case, as explained in \cite{mishra2025graphexgraphbasedextractionmethod}, CTR-based signals provide reliable positive sequence pairs, but not reliable negative pairs. Therefore, we used the MNR loss on the CTR-based labels.\\

\begin{equation}
    \mathcal{L}_{\text{MNR}} = -\log \frac{\exp\left(\frac{\mathbf{z}_i \cdot \mathbf{z}_j}{\tau}\right)}{\sum_{k=1}^{K} \exp\left(\frac{\mathbf{z}_i \cdot \mathbf{z}_k}{\tau}\right)}
\end{equation}

where:

\begin{itemize}
    \item \( \mathbf{z}_i \) and \( \mathbf{z}_j \) are the embeddings of the positive pair,
    \item \( \mathbf{z}_k \) is the embedding of a negative sample,
    \item \( \tau \) is the temperature parameter,
    \item \( K \) is the total number of negative samples.
\end{itemize}

\paragraph{Contrastive Loss}
Contrastive loss \cite{contrastive} explicitly optimizes the embedding space by bringing similar sentence pairs closer together and pushing dissimilar pairs apart. This loss function is therefore well-suited to cases like ours, that rely on Approximate Nearest Neighbor search at prediction time. We used this loss function on both our LLM labels and our SR labels (which both include  positive and negative examples). Mathematically, this loss is defined as:

\begin{equation}
\begin{split}
\mathcal{L}_{\text{con}}
= \tfrac{1}{2}\Bigl(
  & y\, d(u,v)^2 \\
  & + (1-y)\,\max\!\bigl(0,\, m - d(u,v)\bigr)^2
\Bigr)
\end{split}
\end{equation}

where:

\begin{itemize}
    \item \( y \) is a binary label: \( y = 1 \) if the pair is similar, and \( y = 0 \) if the pair is dissimilar.
    \item \( d(u, v) \) is a distance function (cosine distance in our case).
    \item \( m \) is a margin hyperparameter that sets the minimum required separation for dissimilar pairs.
\end{itemize}

This loss function encourages smaller distances for similar sentence pairs (y=1) and larger distances for dissimilar pairs (y=0).\\

\paragraph{Pearson correlation Loss} 
As shown in \cite{liao2024d2llmdecomposeddistilledlarge}, maximizing the Pearson correlation between the student's logits and the teacher's logits enables the student model to replicate the teacher's subtle ranking nuances. It does that by minimizing the Pearson rank imitation loss.

\begin{equation}
\mathcal{L}_{\mathrm{Pearson}}
= 1 - r,
\end{equation}

where the Pearson correlation coefficient $r$ between
the predicted similarity scores $\hat{s}_i$ and target
scores $y_i$ is defined as
\begin{equation}
r =
\frac{
    \sum_{i=1}^{N} 
    \bigl(\hat{s}_i - \bar{\hat{s}}\bigr)
    \bigl(y_i - \bar{y}\bigr)
}{
    \sqrt{
        \sum_{i=1}^{N}
        \bigl(\hat{s}_i - \bar{\hat{s}}\bigr)^2
    }
    \sqrt{
        \sum_{i=1}^{N}
        \bigl(y_i - \bar{y}\bigr)^2
    } + \varepsilon
},
\end{equation}
with $\bar{\hat{s}}$ and $\bar{y}$ denoting the mean values of the
predicted and target similarities, respectively, and
$\varepsilon$ being a small constant to prevent division by~zero.

The Pearson correlation loss encourages a strong linear correlation between the model-predicted similarities and the true similarity labels.

Given two sets of values:
\begin{itemize}
    \item $\hat{s}_i = \tfrac{1}{2}\bigl(\cos(\mathbf{u}_i, \mathbf{v}_i) + 1\bigr)$:
          the rescaled cosine similarity between sentence embeddings
          $\mathbf{u}_i$ and $\mathbf{v}_i$,
    \item $y_i$: the target similarity score in $[0,1]$,
\end{itemize}
\noindent
Minimizing $\mathcal{L}_{\mathrm{Pearson}}$ therefore maximizes the Pearson correlation between predicted and target similarity scores, driving the model to produce embedding-based similarities that align linearly
with the labels.

\paragraph{CoSENT Loss (Cosine Sentence Loss)} This is another loss used during knowledge distillation \cite{Cosent}. Mathematically, it is computed as:

\[
\mathcal{L}_\text{CoSENT} = \log \sum_{(i,j), (k,l)} \left( 1 + \exp(s(i,j) - s(k,l)) \right)
\]

Here, \( (i,j) \) and \( (k,l) \) are any input pairs in the batch such that the cross-encoder-based similarity of \( (i,j) \) is greater than \( (k,l) \). s is the bi-encoder-based similarity function.\\

\paragraph{MSE Loss}
This is the traditional MSE loss, calculated as the Mean Squared Error between the cross-encoder similarity scores, and the cosine similarity scores for the bi-encoder embeddings for items and keyphrases.

\begin{equation}
    \mathcal{L}_{\text{MSE}} = \frac{1}{N} \sum_{i=1}^{N} (y_i - {cos(u,v)}_i)^2
\end{equation}

where $u$ and $v$ are the embeddings for the item and keyphrase respectively and $y$ is the score of the cross-encoder.

\paragraph{KL-Divergence Loss}

Following work done by \cite{yang2020retrievermerelyapproximatorreader, ren-etal-2021-rocketqav2, tamber2025conventionalcontrastivelearningfalls} to distill cross-encoder teachers to retrievers, we use KL-Divergence Loss as a distillation loss.

\begin{equation}
\mathcal{L}_{\mathrm{KL}} 
= \frac{1}{N} 
\sum_{i=1}^{N} 
\sum_{j=1}^{M}
    y_{ij} \,
    \log\!\left(
        \frac{y_{ij}}{\hat{p}_{ij}}
    \right),
\end{equation}

where

\begin{equation}
\hat{p}_{ij} = 
\frac{
    \max\!\bigl( \cos(\mathbf{u}_i, \mathbf{v}_j), 0 \bigr) + \epsilon
}{
    \sum_{k=1}^{M}
        \max\!\bigl( \cos(\mathbf{u}_i, \mathbf{v}_k), 0 \bigr) + \epsilon
}.
\end{equation}

The KL-Diveregence Loss computes the Kullback--Leibler (KL) divergence between the 
\emph{predicted similarity distribution} and the \emph{target probability distribution}.
For each pair of sentence embeddings $(\mathbf{u}_i, \mathbf{v}_j)$, the cosine similarity
is first computed as:

\begin{equation}
s_{ij} = \frac{\cos(\mathbf{u}_i, \mathbf{v}_j) + 1}{2},
\end{equation}
which rescales the similarity to the range $[0, 1]$. 
The similarities are then normalized to form a valid probability distribution:
\begin{equation}
\hat{p}_{ij} = \frac{s_{ij}}{\sum_{k} s_{ik}}.
\end{equation}

Here, $y_{ij}$ denotes the target probability for pair $(i,j)$, 
and $\epsilon$ is a small constant to prevent numerical instability (e.g., $\log(0)$). 
Finally, the KL divergence is averaged over the batch to obtain the overall loss:
\begin{equation}
\begin{split}
\mathcal{L}_{\mathrm{KL}}
= \frac{1}{N}
\sum_{i=1}^{N}\sum_{j=1}^{M}
y_{ij} \log
\frac{y_{ij}}{
    \frac{\exp(s_{ij})}{
        \sum_{k=1}^{M}\exp(s_{ik}) + \varepsilon
    }
}, \\
\text{where} \quad
s_{ij} = \tfrac{1}{2}\!\bigl(\cos(\mathbf{u}_i,\mathbf{v}_j) + 1\bigr).
\end{split}
\end{equation}

\noindent
In summary:
\begin{itemize}
    \item $\mathbf{u}_i, \mathbf{v}_j$ are sentence embeddings from the model.
    \item $\cos(\mathbf{u}_i, \mathbf{v}_j)$ is the cosine similarity between embeddings.
    \item $s_{ij}$ is the rescaled similarity.
    \item $\hat{p}_{ij}$ is the normalized predicted distribution.
    \item $y_{ij}$ is the target probability distribution (each row sums to 1).
    \item $\varepsilon$ is a small constant to avoid $\log(0)$.
    \item $\mathcal{L}_{\mathrm{KL}}$ is the mean KL divergence over the batch.
\end{itemize}

\paragraph{Pairwise Margin MSE Loss}

Given two input sentences $s_1$ and $s_2$, a Sentence Transformer model 
$f(\cdot)$ encodes them into vector representations:
\begin{equation}
    \mathbf{h}_1 = f(s_1), \qquad
    \mathbf{h}_2 = f(s_2).
\end{equation}

The cosine similarity between the embeddings is computed as
\begin{equation}
    \text{cos}(s_1, s_2)
    = \frac{\mathbf{h}_1 \cdot \mathbf{h}_2}
    {\lVert \mathbf{h}_1 \rVert \, \lVert \mathbf{h}_2 \rVert},
\end{equation}
which lies in the interval $[-1, 1]$.  
To align the prediction with target similarity scores in $[0, 1]$, we
apply a linear scaling:
\begin{equation}
    \hat{y}
    = \frac{\text{cos}(s_1, s_2) + 1}{2}.
\end{equation}

Let $y \in [0,1]$ denote the ground-truth similarity score.
The squared prediction error is
\begin{equation}
    e = (\hat{y} - y)^2.
\end{equation}

To make the loss robust to small deviations, a margin $m > 0$ is introduced. Errors are only penalized when they exceed the margin threshold $m^2$.
Formally, the margin mask is defined as
\begin{equation}
    \mathbb{I}_i =
    \begin{cases}
        1, & \text{if } e_i > m^2, \\
        0, & \text{otherwise}.
    \end{cases}
\end{equation}
The masked per-sample loss becomes
\begin{equation}
    L_i = e_i \cdot \mathbb{I}_i.
\end{equation}

For a batch of $N$ sentence pairs, the final Pairwise Margin MSE Loss is
given by
\begin{equation}
    \mathcal{L}
    = \frac{1}{N}
    \sum_{i=1}^{N}
    \left( (\hat{y}_i - y_i)^2 \cdot
    \mathbb{I}\left[(\hat{y}_i - y_i)^2 > m^2\right] \right).
\end{equation}

This objective encourages the predicted similarity to match the target
similarity while ignoring small permissible deviations (i.e., those
within the margin). It therefore yields a robust similarity learning
objective, focusing the model on correcting only those prediction errors
that exceed the allowed margin. We use a margin of 0.3, achieved empirically. 

\paragraph{Softmax Loss}

We follow the setup used in Sentence-BERT~\cite{reimers2019sentencebertsentenceembeddingsusing}, where
a softmax classifier is trained on top of sentence embeddings to perform
natural language inference (NLI) or other sentence-pair classification tasks.

Given a sentence pair $(s_A, s_B)$, a Sentence Transformer model $f(\cdot)$
encodes each sentence into a fixed-dimensional embedding:
\begin{equation}
    \mathbf{u} = f(s_A) \in \mathbb{R}^d, 
    \qquad
    \mathbf{v} = f(s_B) \in \mathbb{R}^d,
\end{equation}
where $d$ is the sentence embedding dimension.

To capture different aspects of the relationship between the two sentence
embeddings, we construct a feature vector by concatenating several
components:
\begin{itemize}
    \item The embeddings themselves: $\mathbf{u}$ and $\mathbf{v}$.
    \item The element-wise absolute difference: $\lvert \mathbf{u} - \mathbf{v} \rvert$.
    \item The element-wise product: $\mathbf{u} \odot \mathbf{v}$.
\end{itemize}

\begin{figure*}[t]
    \centering
    \includegraphics[width=\linewidth]{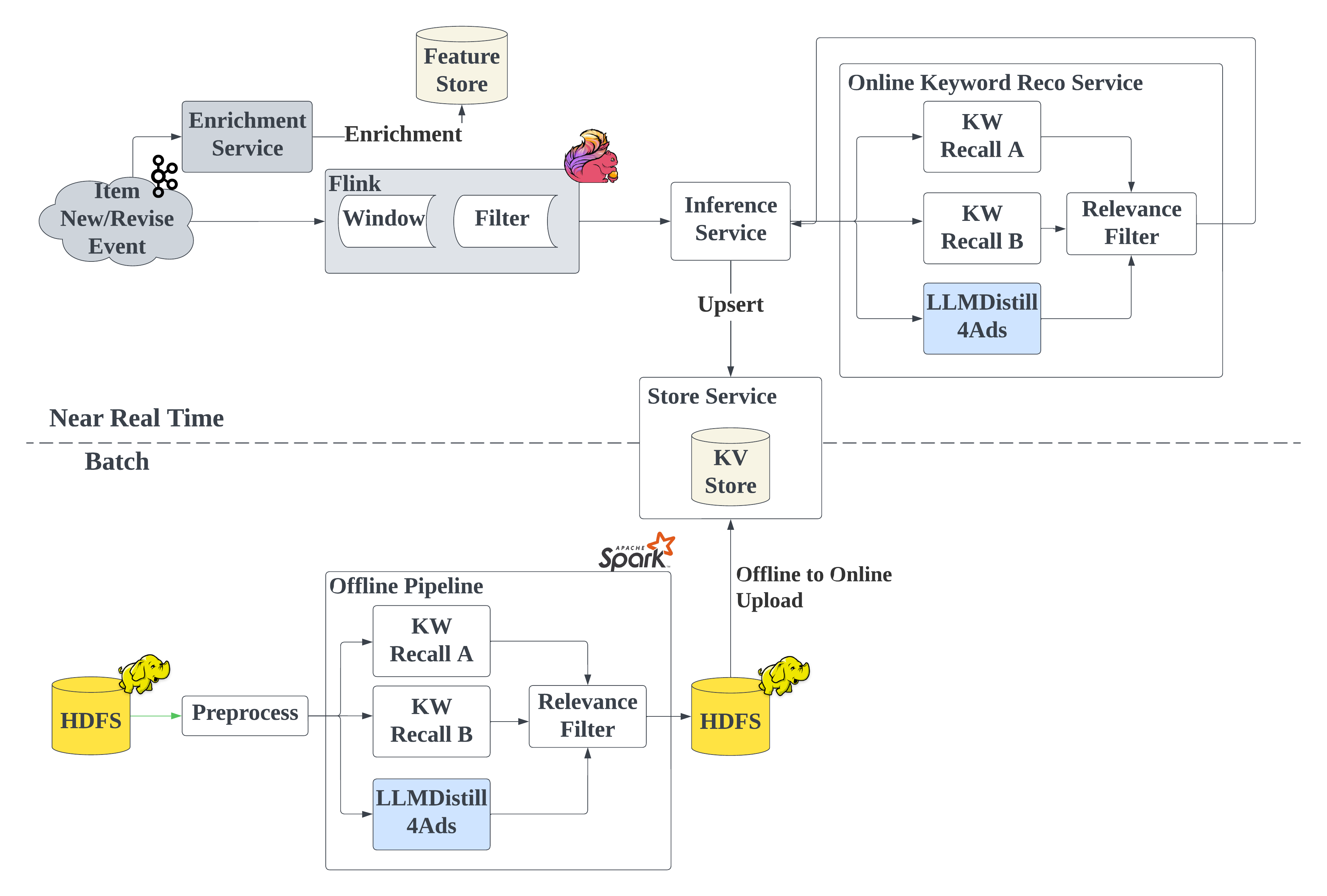}
    \caption{Production Serving Architecture for keyphrase recommendations.}
    \label{fig:prod_arch}
\end{figure*}

In practice, we control which components are used by three boolean
flags:
\begin{itemize}
    \item \texttt{concatenation\_sent\_rep}: include $\mathbf{u}$ and $\mathbf{v}$,
    \item \texttt{concatenation\_sent\_difference}: include $\lvert \mathbf{u} - \mathbf{v} \rvert$,
    \item \texttt{concatenation\_sent\_multiplication}: include $\mathbf{u} \odot \mathbf{v}$.
\end{itemize}

Let $\alpha_{\mathrm{rep}}, \alpha_{\mathrm{diff}}, \alpha_{\mathrm{mult}} \in \{0,1\}$ be
indicator variables specifying whether each component is active. The final
feature vector is then
\begin{equation}
    \mathbf{z}
    =
    \Big[
        \alpha_{\mathrm{rep}} \cdot \mathbf{u}
        \;\Vert\;
        \alpha_{\mathrm{rep}} \cdot \mathbf{v}
        \;\Vert\;
        \alpha_{\mathrm{diff}} \cdot \lvert \mathbf{u} - \mathbf{v} \rvert
        \;\Vert\;
        \alpha_{\mathrm{mult}} \cdot (\mathbf{u} \odot \mathbf{v})
    \Big],
\end{equation}
where $\Vert$ denotes vector concatenation and zeroed components are simply
omitted. The dimension of $\mathbf{z}$ is thus
\begin{equation}
    d_{\mathbf{z}} = d \cdot \big( 2 \alpha_{\mathrm{rep}} + \alpha_{\mathrm{diff}} + \alpha_{\mathrm{mult}} \big).
\end{equation}

On top of the feature vector $\mathbf{z}$, we place a linear classifier
parameterized by a weight matrix $\mathbf{W} \in \mathbb{R}^{C \times d_{\mathbf{z}}}$
and a bias vector $\mathbf{b} \in \mathbb{R}^C$, where $C$ is the number of
class labels. The classifier outputs a logit vector
\begin{equation}
    \boldsymbol{\ell} = \mathbf{W} \mathbf{z} + \mathbf{b}
    \in \mathbb{R}^C.
\end{equation}

The logits are converted into a probability distribution over the $C$
classes via the softmax function:
\begin{equation}
    P(c \mid s_A, s_B) 
    = \frac{\exp(\ell_c)}
           {\sum\limits_{j=1}^{C} \exp(\ell_j)},
    \qquad c = 1, \dots, C.
\end{equation}

Let $y \in \{1, \dots, C\}$ denote the ground-truth class label for the
sentence pair $(s_A, s_B)$. The loss for this example is given by the
standard cross-entropy:
\begin{equation}
    \mathcal{L}
    = - \log P(y \mid s_A, s_B)
    = - \log
    \left(
        \frac{\exp(\ell_y)}
        {\sum\limits_{j=1}^{C} \exp(\ell_j)}
    \right).
\end{equation}

For a batch of $N$ sentence pairs $\{(s_{A,i}, s_{B,i}, y_i)\}_{i=1}^N$,
the overall training objective is the mean cross-entropy loss:
\begin{equation}
    \mathcal{L}_{\text{batch}}
    =
    \frac{1}{N}
    \sum_{i=1}^{N}
    \left[
        - \log
        \left(
            \frac{\exp(\ell_{i, y_i})}
            {\sum\limits_{j=1}^{C} \exp(\ell_{i, j})}
        \right)
    \right],
\end{equation}
where $\boldsymbol{\ell}_i = \mathbf{W} \mathbf{z}_i + \mathbf{b}$ is the
logit vector for the $i$-th sentence pair, and $\ell_{i,j}$ denotes its
$j$-th component.

This Softmax Loss formulation allows the Sentence Transformer to learn
discriminative sentence embeddings tailored for sentence-pair classification tasks such as natural language inference.

\subsection{Other KD Losses}

In pursuing the comparative analysis of MSE (Mean Squared Error, evaluated pointwise), CoSENT, and Pearson Correlation Loss, we explored additional sophisticated ranking loss functions, specifically neural NDCG (Normalized Discounted Cumulative Gain; \citealt{pobrotyn2021neuralndcg}) and Lambda Loss \citep{wang2018lambdaloss}. Unfortunately, the experimental outcomes were significantly suboptimal, with both recall and precision metrics registering values below 0.1. We hypothesize that the substandard results may stem from the typical application of these loss functions, which are generally utilized in conjunction with a seed query and incorporate a penalty for rank misclassification when disparate ranks are involved. This could introduce complications, since the cross-encoder was not specifically trained for precise ranking calibration. Instead, it was designed to optimize overall linear directionality and calibration, which are evidently the primary factors driving its performance. This avenue warrants further investigative research to unravel the underpinning dynamics.

\subsection{Production System Design}
The production architecture depicted in Figure \ref{fig:prod_arch} comprises two main parts: \textit{Near Real-Time (NRT)} Inference and Batch Inference. Batch inference handles items with a delay, while NRT prioritizes immediate items, particularly those newly created or updated by sellers. Batch inference has two components: 1) full batch inference for all items, and 2) daily differential (Diff) to integrate new and updated items with existing data. NRT inference utilizes triton and onnx serving using V100 GPUs, activated by item creation or updates managed by Flink processing and feature enrichment. The full batch handles approximately 2.3 billion items, while the daily Diff supports a churn of 20 million items. As the full batch runs just once, Diff latency determines model deployment viability, being about 35 minutes for bi-encoders. The ANN job downstream takes an additional 2.5 hours daily and for NRT our vector database service helps in that regard. Latency numbers reported for our batch inference use PySpark \cite{spark} (1500 executors, 20g memory, 4 cores), leveraging transformers \cite{wolf-etal-2020-transformers} and onnxruntime \cite{onnxruntime}.
\end{document}